\documentstyle[12pt,aaspp4]{article}
\begin{document}

\title{Gamma-Ray Burst - Supernova Relation}
\footnote{submitted on September 1, 1999 to the proceedings of the Space 
Telescope Science Institute 1999 May Symposium: ``The Largest Explosions Since
the Big Bang: Supernovae and Gamma Ray Bursts"}

\author{Bohdan Paczy\'nski}
\affil{Princeton University Observatory, Princeton, NJ 08544--1001, USA}
\affil{email: bp@astro.princeton.edu}

\begin{abstract}
There is growing evidence that long and hard gamma-ray bursts (GRBs), 
discovered at redshifts between 0.4 and 3.4, are related to some type of 
supernova (SN) explosions.  The GRB ejecta are ultra-relativistic, and
possibly beamed.  There is a possibility that some SN ejecta are
also beamed and/or relativistic.  Prospects for farther advances guided
by expected and unexpected observational developments are very good.
The prospects for developing a sound and quantitative GRB theory any 
time soon are rather modest, if histories of quasars, radio pulsars and
supernovae are used for reference.  However, the current progress in the
understanding of GRB afterglows (which are relativistic) and remnants 
(which are non-relativistic) is likely to continue, as these appear to
be simpler than the GRBs.

According to the current analysis of GRB 970508 the energy of gamma rays
released by this event was about the same as the total energy of explosion.
If correct, this result is difficult to reconcile with the internal shock
models.  It also implies that the global energy generation rate by GRBs is 
four orders of magnitude lower than the rate due to ordinary supernovae, which
makes it very unlikely that the highly energetic supernova remnants were
created by GRBs.
\end{abstract}

%\section 1
\section{Introduction}

The most dramatic recent breakthrough in our understanding of gamma-ray bursts
(GRBs) was made by the BeppoSAX team, which discovered the first X-ray
afterglow (Costa et al. 1997).  That was quickly followed with the discovery
of optical (van Paradijs et al. 1997) and radio (Frail et al. 1997) afterglows,
and the determination of the first optical redshift (Metzger et al. 1997).
By now about two dozen afterglows were detected, almost all within fraction of
an arc second of very faint galaxies, with typical R-band magnitudes 24 - 26.
Approximately ten redshift were measured.  Gradually evidence emerged
that GRBs appear to be associated with star forming regions (Paczy\'nski 1998,
Kulkarni et al. 1998, Galama et al. 1998).  In several cases 
a direct association with a supernova (SN) appeared:
GRB 980425 - SN 1998bw (Galama et al. 1998),
GRB 980326 (Bloom et al. 1999, Castro-Tirado \& Gorosabel 
1999), and GRB 970228 (Reichart 1999, Galama et al. 1999).

We should keep it in mind that all this exciting development is for the
long duration GRBs, as these were the only type for which accurate coordinates
became available within hours of the burst.  The rest of this paper is 
about the long gamma-ray bursts only.

Until recently the most popular models of gamma-ray bursts (GRBs) were related
to merging neutron stars, and neutron stars merging with stellar mass black 
holes.  However, these would be located far away from star forming regions,
and far away from parent dwarf galaxies.  This does not seem to be the case
for the location of GRB afterglows, and this is the reason why an association
of bursts with explosions of massive stars became popular.

Throughout this paper I shall adopt popular assumptions and terminology.
The bursts with strong high energy spectra require very large bulk Lorentz
factors, $ \Gamma > 300 $, to reconcile their rapid variability with their
huge luminosities and no evidence for spectral cut-off due to pair creation
(Baring \& Harding 1996).  During its activity GRB's intensity varies
rapidly.  Several seconds or minutes after the beginning of the burst
an afterglow becomes dominant, as recently shown by Burenin et al. (1999).
The afterglows fade smoothly, usually as a broken power law of time, and 
they are almost certainly due to the interaction between the relativistic
ejecta and ambient medium.  Their emission is non-thermal, and thus it is
fundamentally different form a thermal emission of a non-relativistic 
supernova.  When the ejecta decelerate to non-relativistic
expansion a GRB remnant is created, and at this stage it may resemble a
supernova remnant.  

%\section 2
\section{Rates}

I adopt Hubble constant $ H_0 = 70 ~ km ~ s^{-1} ~ Mpc^{-1} $ throughout
this paper.

According to Wijers et al. (1998) the energy generation rate due to GRBs
is at present epoch (i.e. z = 0) equal 
$$ 
\epsilon _{GRB,0} \approx 10^{52} ~ erg ~ Gpc^{-3} ~ yr^{-1}  , \eqno(1)
$$
assuming that the GRB rate follows the star formation rate as a function 
of redshift.  Note, that this number is independent of beaming of GRB
emission.  If there is beaming the energy per GRB is reduced, but the
the number of GRB explosions increases, so that the product, i.e. 
$ \epsilon _{GRB,0} $ remains unchanged.  Using a very different analysis
Schmidt (1999) obtained the GRB energy generation rate about the same as
Wijers et al. (1998).

The rate of all types of supernovae is approximately
1.5 per $ 10^{10} ~ L_{B, \odot } $ per century (van den Bergh \& 
Tammann 1991).  The mass density of the universe is probably $ \Omega _m
\approx 0.25 $, and the average mass to blue light ratio is $ M/L_B
\approx 200 ~ M_{\odot}/L_{\odot} $ (Bahcall et al. 1995).  Therefore, 
the blue luminosity within one cubic gigaparsec is $ \sim 1.6 \times 10^{17} ~
L_{\odot} $, and the local supernova rate is
$$
n_{SN} \approx 2.4 \times 10^5 ~ Gpc^{-3} ~ yr^{-1} . \eqno(2)
$$
Adopting $ 10^{51} $ erg of kinetic energy per supernova we obtain the
overall energy generation rate (at z = 0)
$$
\epsilon _{SN,0} \approx 2.4 \times 10^{56} ~ erg ~ Gpc^{-3} ~ 
yr^{-1} .  \eqno(3)
$$

It appears that global energy release rate is more than 4 orders of 
magnitude higher for supernovae than it is for gamma-ray bursts 
(Wijers et al. 1998, Schmidt 1999).  Obviously, 
both rates are uncertain.  It is possible that kinetic energy of GRB ejecta
is considerably higher than their gamma ray output (Wijers et al. 1998
Kumar 1999).  It is also possible
that the actual supernova rate is much higher, as intrinsically faint
explosions, like SN 1987A, are difficult to discover, yet they release
about as much energy as ordinary SN Ia or SN II.  While both, $ \epsilon 
_{GRB} $ and $ \epsilon _{SN} $, may well be higher than the estimates
given with the eqs. (1) and (3), it is likely that the ratio
$ \epsilon _{SN} / \epsilon _{GRB} \gg 1 $.  If this seemingly obvious
conclusion is correct it has consequences for finding GRB remnants.

There is no generally accepted quantitative model of GRB emission at this 
time, and we may only guess what is the ratio of gamma-ray energy to kinetic
energy of the ejecta.  While it is common to think that this ratio
is small (Wijers et al. 1998, Kumar 1999), it may just as well be much
larger than unity, i.e. the kinetic energy may turn out to be much
smaller than gamma-ray energy.  This possibility follows from the recent
analysis of the non-relativistic radio remnant of GRB 990508 by Frail,
Waxman and Kulkarni (1999), who find that the total energy is only
$ 5 \times 10^{50} ~ erg $.  At the same time Rhoads (1999b) finds that
GRB 970508 was not strongly beamed, as its afterglow had un-broken power law
decline for over 100 days.  The total gamma-ray emission was at least
$ 3 \times 10^{50} ~ erg $ for this burst (Rhoads 1999b).  If these
claims are correct then for this burst gamma-ray and kinetic energies
were comparable, and this rules out the popular `internal shock' models,
which are very inefficient in generating gamma-rays (e.g. Kumar 1999).

Of course, GRB 970508 was not a typical gamma-ray burst.  Its afterglow
was the only one which first increased in luminosity for about 2 days,
and later declined as un-broken power law for over 100 days.  This is also
the only event for which quantitative estimates were
made for both: gamma-ray and kinetic energies.  We have no direct information
for the ratio of these two energy forms for any other burst.

%section 3
\section{GRB and SN Remnants}

The global energetics of supernovae and gamma-ray bursts has direct
implications for the extra energetic supernova remnants.  Recently,
several suggestions were made that these may be remnants of gamma-ray
burst explosions (Efremov et al. 1998, Loeb \& Perna 1998, Wang 1999).
However, if a typical GRB generates a factor $ f $ more energy than a typical
supernova then the GRB rate must be lower than the supernova rate by a
factor $ 10^4 f $, and correspondingly the number of GRB remnants must
be vastly smaller than suggested by the number of very energetic remnants.
Therefore, it is unlikely that the very energetic supernova remnants
are related to gamma-ray bursts, unless GRBs generate vastly more kinetic
energy than gamma-ray energy.

Let us suppose that the energetic remnants were caused by single explosions.
We know that some rare supernovae are much more powerful than average.
For example, SN 1998bw has released $ \sim 20 \times 10^{51} ~ erg $ (cf.
Woosley et al. 1999, Iwamoto 1999, and references therein, but a much less
energetic explosion has been proposed by Hoflich et al. 1999).  It may well
be that some stellar explosions are even more powerful than SN 1998bw.
However, there is no obvious reason why the most powerful stellar explosions
should be related to gamma-ray bursts.
A classical GRB with a hard spectrum requires
ejecta with the bulk Lorentz factor $ \sim 300 $, or more.  Nobody knows
how to generate outflow so highly ultra-relativistic, and it is not clear
that the total energetics of the explosion has to be extraordinarily large,
as a strongly beamed explosion may appear to be much more energetic that
it really is.  In other words: the ability to generate hard gamma-ray emission
and the overall energetics of an explosion may be correlated with each
other, or just as well the two may be uncorrelated.  As long as we do not have
a sound quantitative model there is no justification for either assumption;
a semi-empirical approach may be more promising than theoretical speculations.

%section 4
\section{GRB and SN Beaming}

The possibility that highly relativistic GRB explosions may be jet-like
was considered for a very long time and I do not know who was the first
to make a suggestion.  Some similarity between the GRBs and
the blazars is so striking that a term `micro-quasar' was suggested some
years ago (Paczy\'nski 1993).  Similarities of these two classes of objects
were recently analyzed by Dermer \& Chiang (1998).  If these are taken
seriously a very strong beaming of GRBs follows, with a drastic reduction 
of the energetics compared to a spherical explosion.  Recently, the breaks
in the rate of decline of several afterglows were interpreted as evidence
for beaming (Kulkarni et al. 1999, Stanek et al. 1999, Harrison et al. 1999).
If GRB emission is confined to a very narrow beam they may not need much
more energy than the `standard' $ \sim 10^{51} ~ erg $ of an
ordinary supernova.  At this time there is no robust estimate
of the degree and the possible range of GRB beaming (e.g. Rhoads 1997, 1999a,b).

More than a decade ago observations of a `mystery spot' near SN 1987A 
were reported by Karovska et al. (1987) and Matcher et al. (1987).  Piran
\& Nakamura (1987) suggested that this might have been a jet generated
by the supernova.  Not knowing about SN 1987A Cen (1998) suggested that
supernovae may create relativistic jets, which may give rise to
gamma-ray bursts.  This idea gained some support when the new analysis of
SN 1987A data provided stronger evidence for the original `mystery spot',
and in addition provided evidence for a second spot on the opposite side
of the supernova, suggesting relativistic jets (Nisenson \& Papaliolios
1999).  Evidence for a strong non-sphericity of SN 1998S was reported
by Leonard et al. (1999).  Hoflich et al. (1999) claim that the explosion
of SN 1998bw was highly non-spherical.

Jets in supernovae became popular (e.g. Khokhlov et al. 1999, Cen 1999,
Nagataki 1999), and often suggested to be associated
with a beamed gamma-ray emission.  A schematic picture may involve
a quasi-spherical and non-relativistic supernova explosion with a 
narrow ultra-relativistic jet streaming along the rotation axis.

The possibility that some supernovae may generate jets is very interesting,
and it should be possible to test it with the VLBA observations of very
young radio supernovae.  However, there is no reason to expect that all
jets must generate gamma-ray bursts, as this would require all outflows to
reach the huge Lorentz factor $ \Gamma \sim 300 $.  It seems much more likely
that there is a broad range of jet velocities, and only some are capable
of GRB-like emission.

%section 5
\section{Hypernova}

While the term `hypernova' became popular recently, it was been sporadically
used in the past (e.g. Wilkinson \& Bruyn 1990).
It does not have a clear, universally accepted meaning.  The following
are several examples.

1. Hypernova is just a name.  The optical light curve of GRB 970508 was 
several hundred times brighter than any SN ever discovered.  The absolute
luminosity of several other afterglows, e.g. 990123, 971214, 990510, was
higher by another factor $ \sim 100 $ (cf. Norris et al. 1999).
So, rather than call it a super-super-nova, or a super-duper-nova, a term
hypernova seems reasonable as a description of the phenomenon, with no
implications for its nature.

2. Hypernova is a special type of a supernova explosion.  At least some
optical afterglows appear to be associated with star forming
regions.  Note that GRBs are many orders of magnitude less common than 
supernovae, and there may be an almost
continuous transition from a typical massive SN to a typical GRB;
the SN 1987A with its relativistic jet and
GRB 980425 - SN 1998bw may be examples of intermediate explosions.
The link between the GRBs and the deaths of
massive stars does not specify the mechanism for a GRB, and
it is testable without a need for theoretical models.  A question: `are GRBs
in star forming regions?' can be answered observationally.
In this context a `hypernova' is an explosion of a massive star, soon
after its formation.  Soon, means several million years, not a delayed
explosion of the merging neutron star type.

3. Hypernova is a rotationally driven supernova.  The idea that at least
some supernovae explosions are driven by a rapid rotation of a compact
core has been around for several decades (e.g. Ostriker \& Gunn 1971).
A qualitative reasoning proceeds as follows.  A spherical collapse
of a massive stellar core transforms $ \sim 3 \times 10^{53} ~ erg $
of gravitational energy into thermal energy of a hot neutron star, and
99.7\% of that energy is lost in a powerful neutrino - anti-neutrino burst,
with the remaining $ \sim 10^{51} ~ erg $ used to power a supernova
explosion.  If a pre-collapse core is rapidly rotating, than additional
$ \sim 3 \times 10^{53} ~ erg $ may be stored in the rotation of
the collapsed object.  Some rotational energy is lost in gravitational
radiation, but a large fraction cannot be readily disposed of.  If an
ultra strong magnetic field is generated by the differential rotation then
it may act as the energy transmitter from the spinning
relativistic object to the envelope, powering an explosion, perhaps in
a form of a relativistic jet.  The more rotation there is, the more
jet-like explosion results, and the more relativistic the jet.  This is
just a speculation at this time, recycled in dozens, perhaps hundreds of
theoretical papers, with terms like a `micro-quasar' (Paczy\'nski 1993)
or a `failed supernova' (Woosley 1993) used at least as often as a `hypernova'
(Paczy\'nski 1998).

%section 6
\section{Pessimistic Conclusions}

It is useful to put theoretical work on gamma-ray bursts in a broader 
perspective of other exotic objects and phenomena in order to asses
the prospects for a short term progress.

There is almost universal agreement that GRB emission is non-thermal.
Several important correlations were found for various GRB properties
(e.g. Fenimore et al. 1995, Liang \& Kargatis 1996, Beloborodov et al. 1998,
Stern et al. 1999, Norris et al. 1999), but is not clear how to incorporate
them in a theoretical model.  This is not surprising.
It is very difficult to prove which specific physical processes are 
responsible for the operation of a non-thermal source - consider
current theories of quasars and radio pulsars.  Well into the 
fourth decade of their development, and no serious ambiguity about the relevant
distance scales, there are no generally accepted theories that account
for either quasar or pulsar non-thermal emission.  There is no reason why 
GRBs should be easier to understand.

For several decades there has been a consensus that Type Ia supernovae 
result from explosive carbon burning in white dwarfs close to the 
Chandrasekhar limit, while all other supernovae are related to core
collapse of various massive stars.  However, the detailed physics is so
complicated that there is still no satisfactory and quantitative
model that could describe the propagation of the nuclear burning front
in SN Ia, without introducing free, adjustable parameters.  There is
also no agreement how $ \sim 0.3\% $ of energy released in core collapse
is channeled to drive the explosion of a SN II.  As far as I can tell,
if there were no observations of SN II it would be impossible to predict
them from the first principles, even though hundreds of sophisticated
papers were written on the subjects.  The guidance provided by the
observations of GRBs and their afterglows is less clear than
it has been for supernovae.  In my view there is no way to prove with
theoretical models that
either merging neutron stars or hypernova explosions should generate
gamma-ray bursts.  It is hard to believe that the puzzle of the central
engine can be solved for GRBs more readily than for supernovae.

There is plenty of observational evidence that a huge diversity of
rotating objects generates either bipolar outflows or jets - the phenomenon
is obviously natural, as it appears so commonly in nature.  Yet, there 
is no quantitative theory of the phenomenon that could explain (without
ad hoc assumptions and ad hoc free parameters or free functions) what
outflow velocities, or what rates of mass loss, should be associated
with any particular object.  The same applies to gamma-ray bursts and
the current attempts to explain why their ejecta are likely to be beamed.

There is no theory that could predict the outflow velocity of any jet,
but it seems natural to expect that only very specific conditions make
it possible to reach the outflow with the Lorentz factor $  \Gamma \sim 300 $,
as needed for HE bursts.  There may be many more jets with more modest
values of $ \Gamma \sim 30, ~ 3 $, or non-relativistic at all.  There is
no direct evidence for a large Lorentz factor for the NHE bursts, which
appear to have no photons above $ \sim 300 ~ keV $ (Pendleton et al. 1997),
and the pair creation
argument does not apply to them.  Perhaps the NHE GRBs are driven by 
non-relativistic explosions.

%section 7
\section{Optimistic Conclusions}

In spite of all theoretical problems there was a spectacular progress in
our understanding of gamma-ray bursts.  The statistics of GRB distribution
obtained with BATSE on Compton Gamma Ray Observatory (Meegan et al. 1992,
Paczy\'nski 1995, and references therein) provided a very strong argument
for a cosmological distance scale to the majority of GRBs.  The obviously
explosive nature of gamma-ray bursts provided the basis for the theoretical
prediction of the afterglows as the products of interaction between GRB
ejecta and ambient medium (Paczy\'nski \& Rhoads 1993, Katz 1994, M\'esz\'aros
\& Rees (1997).  This prediction was confirmed with the discovery
of afterglows with BeppoSAX (Costa et al. 1997), and soon provided the
proof for the cosmological distance (Metzger et al. 1997).  The observed
distribution of the afterglows with respect to host galaxies 
indicated that GRBs are associated with star forming regions, and therefore
with the explosions of massive stars, rather than with merging neutron
stars (Paczy\'nski 1998, Kulkarni et al. 1998, Galama et al. 1998).  There
is evidence that at least some bursts are directly associated with explosions
of some supernovae (Galama et al. 1998, Bloom et al. 1999, Castro-Tirado \& 
Gorosabel 1999, Reichart 1999, Galama et al. 1999).

There is every reason to expect more progress along similar lines: observations
and their analysis providing more and more hints about the nature of the
bursts.  The following are some of the likely lines of progress in our
understanding.

The new GRB instruments will provide hundreds of accurate positions within
seconds of the burst's beginning, for long as well as for short bursts.  We
may expect that the distribution in distance will soon be known not only
for the long HE bursts, but also for the NHE bursts and for the short bursts.
It may well be that in several years some GRB will be the redshift record
holder.  If GRBs trace massive star formation rate, then they may become
a new probe of the process in very dusty regions, or at very high redshifts.

While old GRB remnants may be difficult to distinguish from SN remnants,
there is a possibility that a clear signature of the effect of non-thermal
emission from a GRB and its afterglow may be detected in the interstellar
medium (e.g Perna et al. 1999, Draine 1999, Weth et al. 1999), and it may turn
out to be a powerful new diagnostics for these events.  The importance if
the interstellar scintillation for the estimates of radio afterglow expansion
has already proven to be an important research tool (Goodman 1997, Frail et 
al. 1997).

If GRBs are related to explosions of massive stars then we expect
that a circum-stellar gas is a leftover from a strong stellar wind,
as all massive stars appear to have winds.  Currently there is mixed
evidence from afterglow studies, with some events consistent with
ambient gas density falling off as $ 1/r^2 $, as expected of wind
environment, while in some the ambient gas density appeared to be
constant (Chevalier \& Li 1999a,b).  With many more afterglows followed
with multi-band studies it will be possible to determine which 
environment is more common, and to make inferences about the nature of
the exploding object.

At a cost much lower than any GRB space mission a super-super ROTSE or a
super-super-LOTIS may be developed to follow up on the experience of
ROTSE (Akerlof et al. 1999) and LOTIS (Williams et al. 1999).  At a cost
less than $ \$10^6 $ it should be possible to implement an all sky optical
monitoring system sensitive to optical flashes of $ \sim 1 $ minute
duration, like the one discovered by ROTSE (Akerlof et al. 1999), 
detectable without
any GRB trigger.  There may be many more optical flashes than gamma-ray
bursts if less extreme Lorentz factors are sufficient for generating
optical flashes.  Rather obviously, a major difficulty is not hardware 
but software.

We already know that some supernovae (SN 1998bw) eject some matter
at a relativistic or sub-relativistic velocity (Waxman \& Loeb 1999).
There is a fairly strong case for a relativistic jet from the SN 1987A
(Nisenson \& Papaliolios 1999).  We may expect (or at least hope) that
other cases of relativistic motion will be discovered in other SN.  For 
supernovae within $ \sim 100 $ Mpc it may be possible to detect
anisotropy in their ejecta, perhaps even superluminal jets, using VLBA.
If jets are detected in many cases it will be possible to study the
distribution of jet velocities.

When the number of recorded supernova explosions will exceed $ 10^4 $ we
shall know more about the high energy tail of their power distribution,
and we may learn if there is a sharp maximum, or is there an extended
tail, to the explosions in the $ 10^{53} - 10^{54} ~ erg $ range.

The ever more vigorous searches for distant (i.e. faint) supernovae will
discover optical afterglows without a need for the GRB alert (Rhoads 1997).
There may be a rich diversity of SN-like or afterglow-like events, perhaps
even optical transients from merging neutron stars (Li \& Paczy\'nski 1998).

If the past can be used as a guide for the future than the most spectacular
breakthroughs in the observations and understanding of gamma-ray bursts
will be unexpected, just as the most recent BeppoSAX breakthrough was.
An example may be the recent empirical finding of a very tight correlation
between photon energy-dependent lags and peak luminosities of gamma-ray bursts
(Norris et al. 1999).

\vskip 0.5cm

This work was not supported by any grant.

%REFERENCES

\end{document}